\documentclass[aps,pra,showpacs,twocolumn]{revtex4}
\newcommand{\be}{\begin{equation}}
\newcommand{\ee}{\end{equation}}
\newcommand{\brr}{\begin{eqnarray}}
\newcommand{\err}{\end{eqnarray}}
\newcommand{\nn}{\nonumber}
\newcommand{\bd}{\begin{displaymath}}
\newcommand{\ed}{\end{displaymath}}

\newcommand{\bib}{\bibitem}
\newcommand{\bfig}{\begin{figure}}
\newcommand{\efig}{\end{figure}}
\newcommand{\ie}{i.e.}

\def\lam{\lambda}

\def\eps{\epsilon}
\def\rpar{\right)}
\def\lpar{\left(}
\def\rbk{\right]}
\def\lbk{\left[}
\def\rbr{\right\}}
\def\lbr{\left\{}
\def\lb{\label}
\def\ro{\mbox{\boldmath $\rho$}}
\def\opi{\mbox{\boldmath $\Pi$}}
\def\rg{\rangle}

\def\nc{\mathrm{i}}
\def\coloneq{\mathrel{\mathop:}=}
\def\half{\frac{1}{2}}

\def\fa{\mathfrak{a}}
\def\fq{\mathfrak{q}}
\def\fz{\mathfrak{z}}
\usepackage{amssymb}
\usepackage{mathrsfs}
\usepackage{upmath}
\usepackage{yfonts}
\usepackage{dsfont}
\usepackage{amsmath}
\usepackage{graphicx}
\usepackage{bm}
\usepackage{amsfonts}
\usepackage{eucal}
\begin{document}
\title{Quasiprobability distribution functions for finite-dimensional discrete phase spaces: spin tunneling effects in a toy model}
\author{Marcelo A. Marchiolli}\email{mamarchi@ift.unesp.br}
\author{Evandro C. Silva}\email{mandrake@ift.unesp.br}
\author{Di\'{o}genes Galetti}\email{galetti@ift.unesp.br}
\affiliation{Instituto de F\'{\i}sica Te\'{o}rica, Universidade Estadual Paulista, Rua Pamplona 145, 01405-900, S\~{a}o Paulo, SP, Brazil}
\date{\today}
\begin{abstract}
We show how quasiprobability distribution functions defined over $N^{2}$-dimensional discrete phase spaces can be used to treat physical 
systems described by a finite space of states which exhibit spin tunneling effects. This particular approach is then applied to the
Lipkin-Meshkov-Glick model in order to obtain the time evolution of the discrete Husimi function, and as a by-product the energy gap for 
a symmetric combination of ground and first excited states. Moreover, we also show how an angle-based potential approach can be efficiently 
employed to explain qualitatively certain features of the energy gap in terms of a spin tunneling. Entropy functionals are also discussed 
in this context. Such results reinforce not only the formalism {\it per se} but also the possibility of some future potential applications 
in other branches of physics.
\end{abstract}
\pacs{03.65.Ca, 03.65.Xp, 21.60.Fw} 
\maketitle
\section{Introduction}

In the last decades, much effort has been devoted to characterize quantum tunneling processes in mesoscopic and/or macroscopic systems, 
emphasizing the importance of the degree of freedom related to the angular momentum and angle pair \cite{Takagi}. In what concerns the 
spin tunneling, certain theoretical approaches have pointed to some different ways of treating this problem, each one presenting a 
particular set of convenient inherent mathematical properties \cite{Enz}. From this perspective, if one considers physical systems with a 
finite-dimensional space of states and described by discrete variables, a sound theoretical framework must be employed to characterize 
properly such nonclassical effect. In fact, an alternative approach to the system description of these specific cases can be pointed out. 
First, we recognize that the state spaces associated with those particular physical systems are $N$-dimensional Hilbert spaces. Next,
in connection with these finite Hilbert spaces, it should be stressed that quantum representations of $N^{2}$-dimensional discrete phase 
spaces can also be constructed \cite{Galetti1}. Thus, relevant operators whose kinematical and/or dynamical contents carry all the 
necessary information for describing those quantum systems can now be promptly mapped in such phase spaces. In this sense, although there 
are various phase spaces formalisms proposed in the literature for treating finite-dimensional physical systems \cite{Vourdas}, let us focus 
our attention upon the framework developed in Refs. \cite{Marcelo1,Ruzzi1,Ruzzi2,Marcelo2,Marcelo3} for the discrete representatives of the
quasiprobability distribution functions defined in $N^{2}$-dimensional phase spaces, which has its algebraic structure based on the 
technique of constructing unitary operator bases initially formulated by Schwinger \cite{Schwinger}. The virtue of this discrete quantum
phase-space approach is that it allows us to exhibit and handle the pair of complementary variables related to a particular degree of
freedom we are dealing with, as well as to recognize the quantum correlations between them. The basic idea then consists in exploring this
mathematical tool in order to study those quantum correlations in connection with spin tunneling processes. Therefore, besides having the
energy spectrum, which can be obtained via direct diagonalization of the Hamiltonian system, we can also have additional quantum 
information about the physical system through the study of the corresponding discrete Wigner and/or Husimi functions.

Such theoretical framework is then applied, in particular, to the Lipkin-Meshkov-Glick (LMG) model \cite{Lipkin}, which was originally 
introduced over forty years ago in nuclear physics \cite{Ring} for treating certain fermionic systems. This important toy model has, since 
then, been extensively studied in the literature because of its apparent simplicity \cite{Varios}. Indeed, it can also be viewed as a 
finite set of spins half mutually interacting in the $xy$-plane subjected to a transverse magnetic field \cite{Vidal}. Moreover, our 
interest in this model also resides in the fact that spin tunneling can be considered to occur.

In this work, we show how the time-dependent discrete Husimi function and an angle-based potential description \cite{Ruzzi3,Pimentel,Galetti2} 
can be combined in order to describe a group of physical processes that encompasses, among other things, the spin tunneling effects. This 
particular approach leads us not only to extract the energy gap for a symmetric combination of ground and first excited states, but also to
corroborate its inherent applicability to analogous physical systems such as in magnetic molecules \cite{Galetti3}.

This paper is organized as follows. In Section II, we present a condensed review of the theoretical apparatus used to describe 
$N^{2}$-dimensional phase spaces and also discuss some new essential features exhibited by the time-dependent discrete Husimi function.
In Section III, we apply our results to the LMG model to explore qualitatively the spin tunneling effects for a symmetric combination of 
ground and first excited states. Finally, Section IV contains our summary and conclusions.

\section{Theoretical apparatus for finite-dimensional phase spaces}

Our theoretical framework is totally based upon the formalism developed in Refs. \cite{Ruzzi2,Marcelo2} for physical systems with
finite-dimensional space of states. In this sense, let us introduce some basic elements which represent our guidelines for the
fundamentals of the formal description of $N^{2}$-dimensional phase spaces by means of discrete variables. The first important element 
is the mod$(N)$-invariant operator basis
\be
\lb{e1}
{\bf T}^{(s)}(\mu,\nu) = \frac{1}{\sqrt{N}} \!\!\! \sum_{\eta,\xi = - \ell}^{\ell} \exp \lbk - \frac{2 \pi \nc}{N} (\eta \mu + \xi \nu)
\rbk {\bf S}^{(s)}(\eta,\xi) \, ,
\ee
which consists of a discrete Fourier transform of the extended mapping kernel ${\bf S}^{(s)}(\eta,\xi) = [ \mathscr{K}
(\eta,\xi)]^{-s}{\bf S}(\eta,\xi)$,
\be
\lb{e2}
{\bf S}(\eta,\xi) \coloneq \frac{1}{\sqrt{N}} \exp \lpar \frac{\nc \pi}{N} \eta \xi \rpar {\bf U}^{\eta} {\bf V}^{\xi} 
\ee
being the symmetrized version of the unitary operator basis proposed by Schwinger \cite{Schwinger}. In this particular approach, the 
labels $\eta$ and $\xi$ are associated with the dual momentum- and coordinate-like variables of a discrete $N^{2}$-dimensional phase space. 
Note that these labels are congruent modulo $N$ and assume integer values in the symmetrical interval $[-\ell,\ell]$ for $\ell = (N-1)/2$ 
fixed. Besides, the extra term $\mathscr{K}(\eta,\xi)$ is defined through the ratio $\mathscr{M}(\eta,\xi) / \mathscr{M}(0,0)$, where the 
function \cite{Ruzzi2}
\brr
& & \mathscr{M}(\eta,\xi) = \frac{\sqrt{\fa}}{2} \Bigl\{ \vartheta_{3} (\fa \eta | \nc \fa) \Bigl[ \vartheta_{3} (\fa \xi | \nc \fa) 
+ \vartheta_{4} (\fa \xi | \nc \fa) \mathrm{e}^{\nc \pi \eta} \Bigr] \nn \\
& & \quad + \; \vartheta_{4} (\fa \eta | \nc \fa) \Bigl[ \vartheta_{3} (\fa \xi | \nc \fa) \mathrm{e}^{\nc \pi \xi} + \vartheta_{4} 
(\fa \xi | \nc \fa) \mathrm{e}^{\nc \pi (\eta + \xi + N)} \Bigr] \Bigr\} \nn
\err
with $\fa = (2N)^{-1}$, is responsible for the sum of products of Jacobi theta functions evaluated at integer arguments, and $s$ refers
to a complex parameter that satisfies the relation $| s | \leq 1$. It is worth mentioning that a compilation of results and properties 
which characterize the algebraic structure of the discrete mapping kernel ${\bf T}^{(s)}(\mu,\nu)$, as well as the unitary operators 
${\bf U}$ and ${\bf V}$, can be promptly found in Refs. \cite{Marcelo1,Ruzzi2}. For physical applications related to quantum tomography 
and quantum teleportation, see also Ref. \cite{Marcelo2}.

Next, let us assume that $\ro(t)$ reflects the dynamics of a particular quantum system characterized by a finite-dimensional space of
states. The one-to-one mapping between density operators and functions belonging to an $N^{2}$-dimensional phase space labelled by 
$\{ \mu,\nu \} \in [-\ell,\ell]$ is attained, in our description, by means of the parametrized function $F^{(s)}(\mu,\nu;t) \coloneq
\mathrm{Tr} [ {\bf T}^{(s)}(\mu,\nu) \ro(t) ]$. Expressed as a double discrete Fourier transform of the discrete $s$-ordered
characteristic function $\Xi^{(s)}(\eta,\xi;t) \coloneq \mathrm{Tr} [ {\bf S}^{(s)}(\eta,\xi) \ro(t) ]$, it has a well-established
continuous counterpart within the Cahill-Glauber formalism \cite{Glauber}. Moreover, for $s=-1,0,1$ the time-dependent function
$F^{(s)}(\mu,\nu;t)$ is directly related to the respective discrete Husimi, Wigner, and Glauber-Sudarshan distribution functions. An
interesting formal result from this formalism explores the connection between discrete Wigner and Husimi functions -- here denoted by
$\mathcal{W}(\mu,\nu;t)$ and $\mathcal{H}(\mu,\nu;t)$ -- through the equation
\be
\lb{e3}
\mathcal{H}(\mu,\nu;t) = \frac{1}{N} \sum_{\mu^{\prime},\nu^{\prime} = - \ell}^{\ell} \mathrm{E}(\mu,\nu | \mu^{\prime}, \nu^{\prime})
\mathcal{W}(\mu^{\prime},\nu^{\prime};t) \, ,
\ee
where $\mathrm{E}(\mu,\nu | \mu^{\prime}, \nu^{\prime}) \equiv \mathrm{Tr} \lbk {\bf T}^{(0)}(\mu,\nu) {\bf T}^{(-1)}(\mu^{\prime},
\nu^{\prime}) \rbk$ defines a smoothing process characterized by a discrete phase-space function that closely resembles the role of a
Gaussian function in the continuous phase space. In fact, Eq. (\ref{e3}) represents an intermediate smoothing sequence within a 
hierarchical process among the quasiprobability distribution functions in finite-dimensional spaces \cite{Marcelo2}. In what concerns
the time evolution of the density operator, it must be stressed that $\ro(t)$ satisfies the von Neumann-Liouville equation and its
corresponding mapped expression in the discrete phase-space representation leads us to obtain a differential equation for $\mathcal{W}
(\mu,\nu;t)$, whose solution was explicitly determined in Ref. \cite{Ruzzi1}. Hence, the discrete Husimi function can now be immediately 
inferred.  

Now, let us establish an important mathematical result for the time-dependent discrete Husimi functions. First of all, it should be 
noticed that $\mathcal{H}(\mu,\nu;t)$ is strictly positive and limited to the interval $[0,1]$ for any $t \geq 0$; secondly, the phase 
space treated here consists of a finite mesh with $N^{2}$ points and characterized by the discrete variables $\mu$ and $\nu$. So, the 
discrete Husimi function can be mapped onto a real $N \times N$ matrix $\mathds{H}(t)$ whose elements $\lbk h_{rs}(t) \rbk_{r,s = 1, 
\ldots, N}$ obey two essential properties that imply in the conservation of probabilities on a discrete phase space, that is,
\brr
0 \leq h_{rs}(t) \leq 1 \quad \mbox{and} \quad \sum_{r,s=1}^{N} h_{rs}(t) = 1 \, . \nn
\err
The interaction with any dissipative environment is automatically discarded within this context. Such mathematical procedure brings 
some operational advantages in our description since the matrix $\mathds{H}(t)$ can be promptly diagonalized onto the eigenspaces
\bd 
\mathrm{V}_{\lam_{i}} = \{ \mathds{V}_{i} \in \mathrm{V} : \mathds{H} \mathds{V}_{i} = \lam_{i} \mathds{V}_{i} \} 
\ed
characterized by the eigenvectors $\{ \mathds{V}_{i}(t) \}_{i=1,\ldots,N}$ and their corresponding eigenvalues $\{ \lam_{i}(t) \}_{i=1,\ldots,N}$ 
for $t$ fixed. It is worth emphasizing that the eigenvalues obtained from this particular diagonalization process assume, in general,
both real and complex values, and this fact can be explained by means of matrix analysis \cite{Aldrovandi}. A pertinent question
then emerges from our considerations on $N^{2}$-dimensional phase spaces: ``Can both real and complex eigenvalues be associated with
some physical process?"

To answer this question, let us initially decompose the matrix $\mathds{H}(t)$ as a sum of two Hermitian and antihermitian 
$N \times N$ matrices, namely, $\mathds{H}(t) = \mathds{A}(t) + \mathds{B}(t)$, both matrices being constructed out following the
mathematical recipe $\mathds{A}(t) = \half \lbk \mathds{H}(t) + \mathds{H}^{\dagger}(t) \rbk$ and $\mathds{B}(t) = \half \lbk 
\mathds{H}(t) - \mathds{H}^{\dagger}(t) \rbk$. In this situation, the diagonalization process attributes real eigenvalues for 
$\mathds{A}(t)$, while $\mathds{B}(t)$ has eigenvalues which are pure imaginary (or zero). Besides, the trace of $\mathds{H}(t)$ is 
preserved, \ie,
\brr
\mathrm{Tr} \lbk \mathds{H}(t) \rbk \equiv \mathrm{Tr} \lbk \mathds{a}(t) \rbk = \sum_{i=1}^{N} \sigma_{\mathds{A}}^{(i)}(t) \, . \nn
\err
Here, $\{ \sigma_{\mathds{A}}^{(i)}(t) \}_{i=1,\ldots,N}$ and $\{ \sigma_{\mathds{B}}^{(i)}(t) \}_{i=1,\ldots,N}$ represent the 
respective eigenvalues of the matrices $\mathds{A}(t)$ and $\mathds{B}(t)$ for all $t \geq 0$; therefore, it is easy to verify that
these eigenvalues are now responsible for the real and imaginary parts of the complex eigenvalues $\{ \lam_{i}(t) \}_{i=1,\ldots,N}$.
Next, let us introduce an auxiliary tool characterized by the entropy functional
\be
\lb{e4}
\mathrm{E} \lbk \{ \lam_{i} \}; t \rbk = - \sum_{i=1}^{N} | \lam_{i}(t) | \ln \lpar | \lam_{i}(t) | \rpar \, ,
\ee
which allows us to infer all the different contributions associated with $\{ \lam_{i}(t) \} \in \mathbb{C}$. From the operational
point of view, the simplicity of this measure represents an effective gain to our task since the diagonalization process of the
time-dependent discrete Husimi function is sufficient in this case for determining $\mathrm{E} [ \{ \lam_{i} \}; t ]$. The next step
then consists in considering a well-known physical system that leads us to find out any concrete evidence in (\ref{e4}) of some 
particular physical process inherent to the model. In this sense, we will apply the theoretical framework here discussed to a solvable 
quasi-spin model whose Hamiltonian, although simple, presents some interesting physical and mathematical features, namely, the 
Lipkin-Meshkov-Glick (LMG) model \cite{Lipkin}.

\section{The LMG model}

Originally proposed with the intent of testing mean-field approximations in many-body systems, the LMG model is here introduced through
the Hamiltonian \cite{Lipkin}
\brr
{\bf H} = \frac{\eps}{2} \sum_{\fq , \sigma} \sigma \, {\bf a}_{\fq , \sigma}^{\dagger} {\bf a}_{\fq , \sigma} + \frac{V}{2}
\sum_{\fq, \fq^{\prime}, \sigma} {\bf a}_{\fq, \sigma}^{\dagger} {\bf a}_{\fq^{\prime}, \sigma}^{\dagger} 
{\bf a}_{\fq^{\prime},- \sigma} {\bf a}_{\fq,- \sigma} \nn
\err
which describes a collection of $\mathrm{N}_{p}$ fermions distributed in two $\mathrm{N}_{p}$-fold degenerate levels separated by an
energy $\eps$. The degenerate states within each level are labelled in this expression by means of the quantum numbers 
$\fq \in [1, \mathrm{N}_{p}]$ and $\sigma = \pm 1$ ($+1$ and $-1$ represent the respective higher and lower levels), $\mathrm{N}_{p}$
being considered an even number. 

It is worth noticing that the introduction of the quasi-spin operators \cite{Ring}
\brr
{\bf J}_{\pm} \coloneq \sum_{\fq} {\bf a}_{\fq, \pm 1}^{\dagger} {\bf a}_{\fq, \mp 1} \quad \mbox{and} \quad {\bf J}_{\fz} \coloneq
\half \sum_{\fq, \sigma} \sigma \, {\bf a}_{\fq , \sigma}^{\dagger} {\bf a}_{\fq , \sigma} \nn
\err
into the LMG Hamiltonian not only reveals its underlying $SU(2)$ structure, since the operators ${\bf J}_{\pm}$ and ${\bf J}_{\fz}$ 
obey the standard commutation relations $[ {\bf J}_{+},{\bf J}_{-} ] = 2 {\bf J}_{\fz}$ and $[ {\bf J}_{\fz},{\bf J}_{\pm} ] = \pm 
{\bf J}_{\pm}$, but also allows us to treat collective excitations of the fermionic system in a more suitable form, namely 
${\bf H} = \eps {\bf J}_{\fz} + (V/2) \lpar {\bf J}_{+}^{2} + {\bf J}_{-}^{2} \rpar$. Indeed, the ${\bf J}_{\fz}$ term of this
Hamiltonian operator gives half the difference of the number of particles laid on the upper and lower levels, while the second term,
involving the operators ${\bf J}_{+}^{2}$ and ${\bf J}_{-}^{2}$, is associated with the interaction between a pair of particles
located at the same energy level, being also responsible for the scattering process of this pair to the other level where the quantum
number $\fq$ of each particle is preserved. For simplicity, let us rewrite such Hamiltonian operator in order to scale the interaction 
term to the particle number while measuring the energy in terms of $\eps$, \ie,
\be
\lb{e5}
{\bf H}_{L} \coloneq \frac{{\bf H}}{\eps} = {\bf J}_{\fz} + \frac{\chi}{2 \mathrm{N}_{p}} \lpar {\bf J}_{+}^{2} + {\bf J}_{-}^{2} \rpar
\ee
with $\chi = \mathrm{N}_{p} V / \eps$. Since $[ {\bf H}_{L},{\bf J}^{2} ] = 0$, it turns immediate to see that ${\bf H}_{L}$ can be 
diagonalized within each $(2J+1)$-dimensional multiplet labelled by the eigenvalues of ${\bf J}^{2}$ and ${\bf J}_{\fz}$, which 
accounts for the soluble character of the associated quantum model \cite{Lipkin,Ring}. It is worth stressing that the ground state
belongs to the finite multiplet characterized by $\mathrm{J} = \frac{\mathrm{N}_{p}}{2} = \mathrm{max}(\mathrm{J}_{\fz})$, so that its
${\bf J}^{2}$ and ${\bf J}_{\fz}$ quantum numbers are $\lpar \frac{\mathrm{N}_{p}}{2} \rpar \lpar \frac{\mathrm{N}_{p}}{2} + 1 \rpar$
and $- \frac{\mathrm{N}_{p}}{2}$, respectively. Hereafter, we will be interested only in this particular multiplet containing the
ground state and whose dimension is given by $N = 2 \mathrm{J}+1 = \mathrm{N}_{p}+1$. This $(\mathrm{N}_{p}+1)$-dimensional multiplet
will be considered as our underlying Hilbert space of interest \cite{note1}, and also will be used to construct the
$(\mathrm{N}_{p}+1)^{2}$-dimensional discrete phase space.      

Next, let us mention some few words about two discrete conserved quantities inherent to the LMG model which reflect certain symmetry
properties. The simplest operator commuting with the Hamiltonian (\ref{e5}), therefore giving a constant of motion, is the parity 
operator $\opi \coloneq \exp ( \nc \pi {\bf J}_{\fz} )$. This fact tells us that the Hamiltonian matrix, in the ${\bf J}_{\fz}$ 
representation, breaks into two disjoint blocks involving only even and odd eigenvalues of ${\bf J}_{\fz}$, respectively. The second 
interesting quantity comes from the anticommutation relation $\{ {\bf H}_{L},{\bf R} \} = 0$, where  
\brr
{\bf R}(-\pi/2,\pi,0) \coloneq \exp [ \nc (\pi/2) {\bf J}_{\fz} ] \exp ( \nc \pi {\bf J}_{y} ) \nn
\err
corresponds to a rotation of the angular momentum quantization frame by the Euler angles $(-\pi/2,\pi,0)$, thus transforming 
${\bf H}_{L} \rightarrow - {\bf H}_{L}$. In this case, if $| \mathrm{E}_{j} \rg$ is an energy eigenstate with eigenvalue 
$\mathrm{E}_{j}$, then ${\bf R} | \mathrm{E}_{j} \rg$ is also an eigenstate of ${\bf H}_{L}$ with eigenvalue $- \mathrm{E}_{j}$. This
symmetry property of the Hamiltonian operator (\ref{e5}) gives rise to an energy spectrum that is symmetric about zero. 

\begin{figure}[!t]
\centering
\includegraphics[width=5cm]{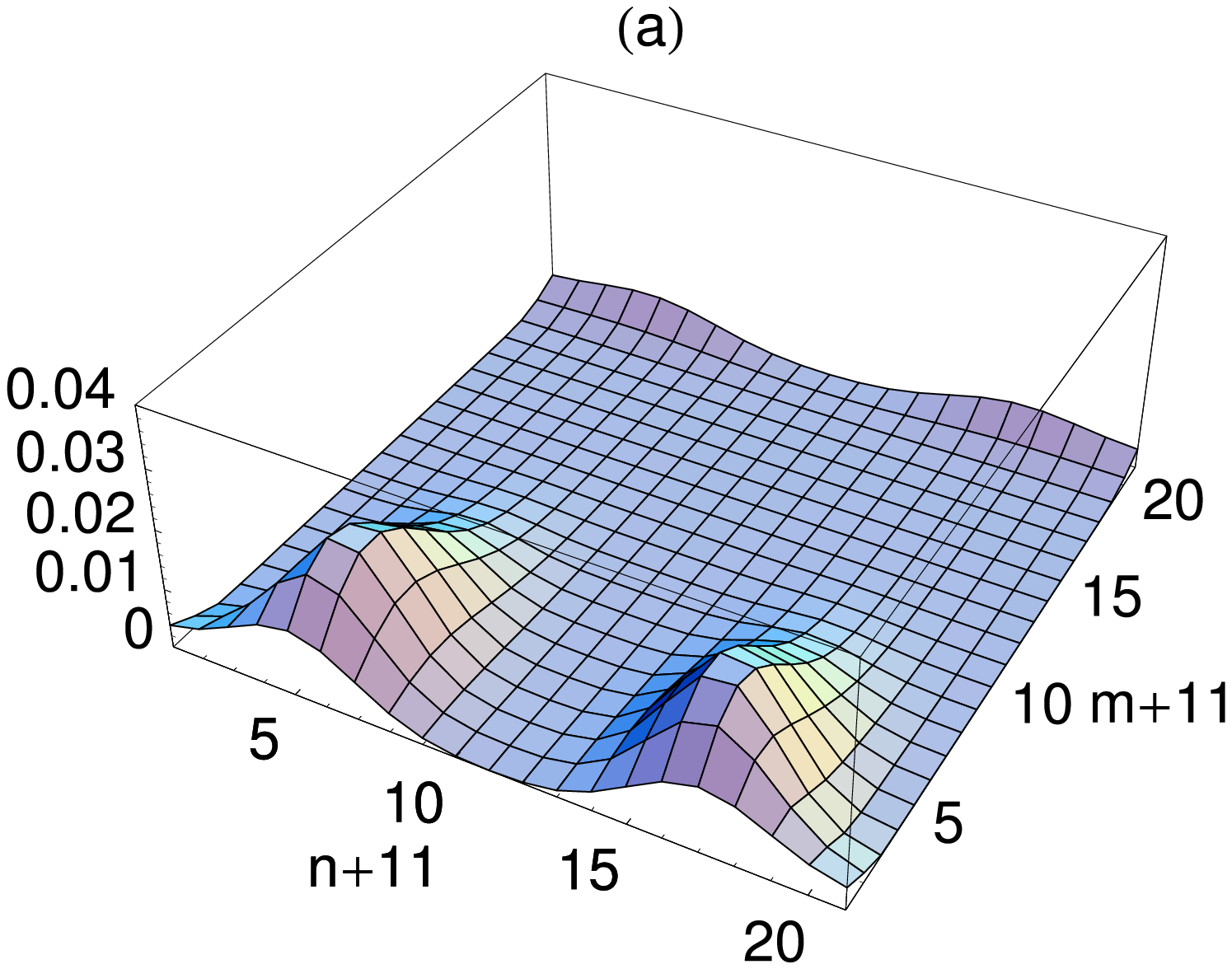}
\includegraphics[width=5cm]{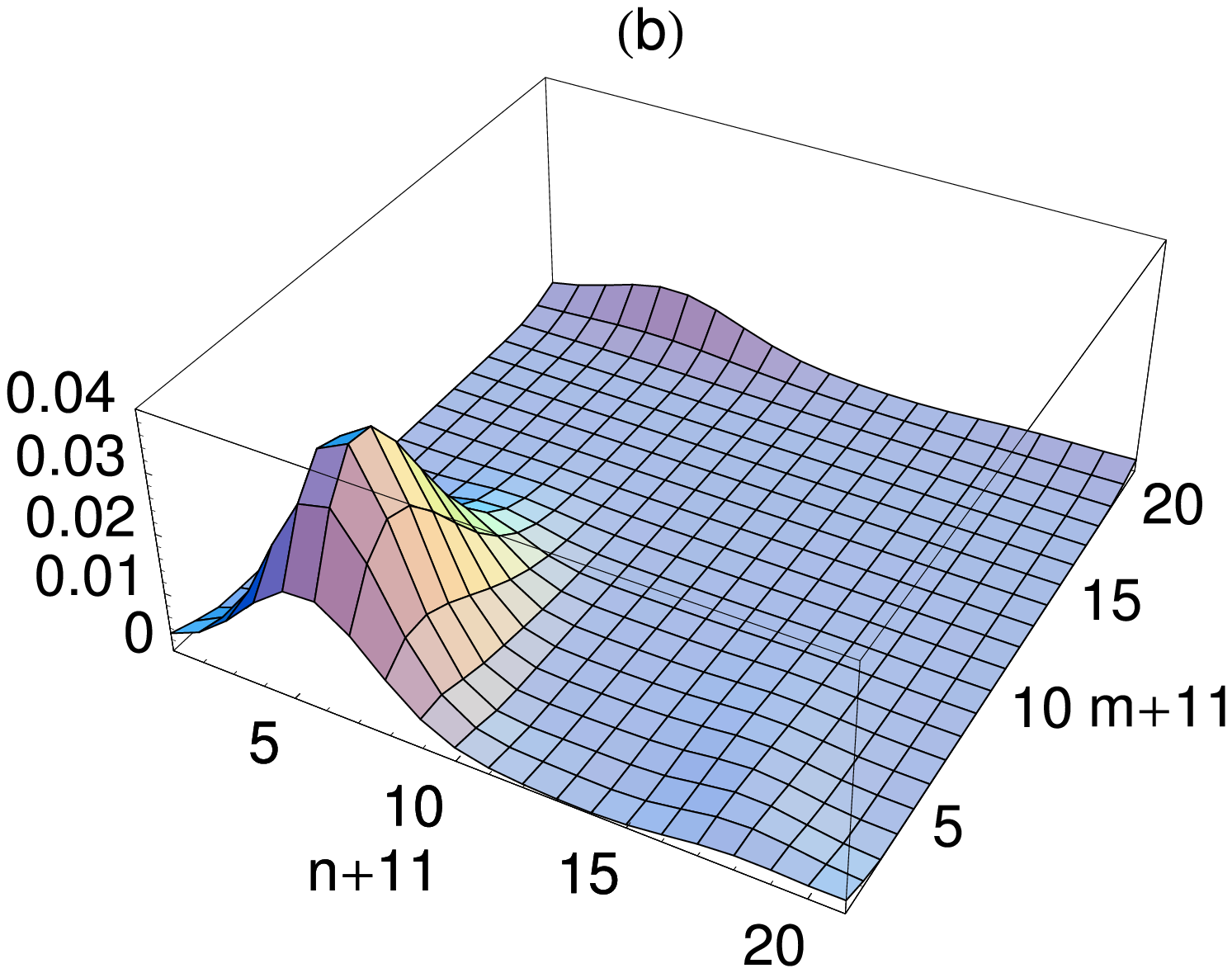}
\includegraphics[width=5cm]{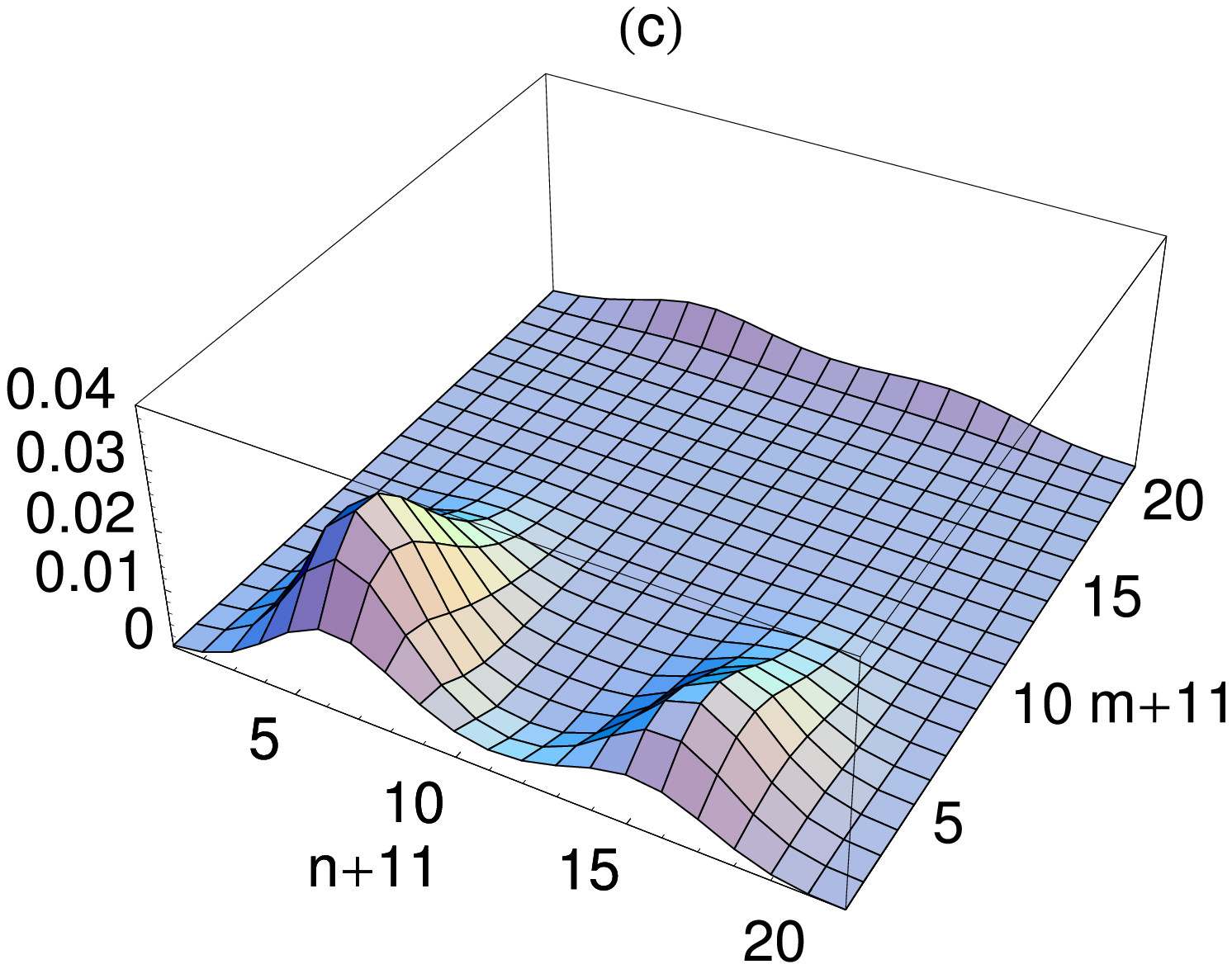}
\includegraphics[width=5cm]{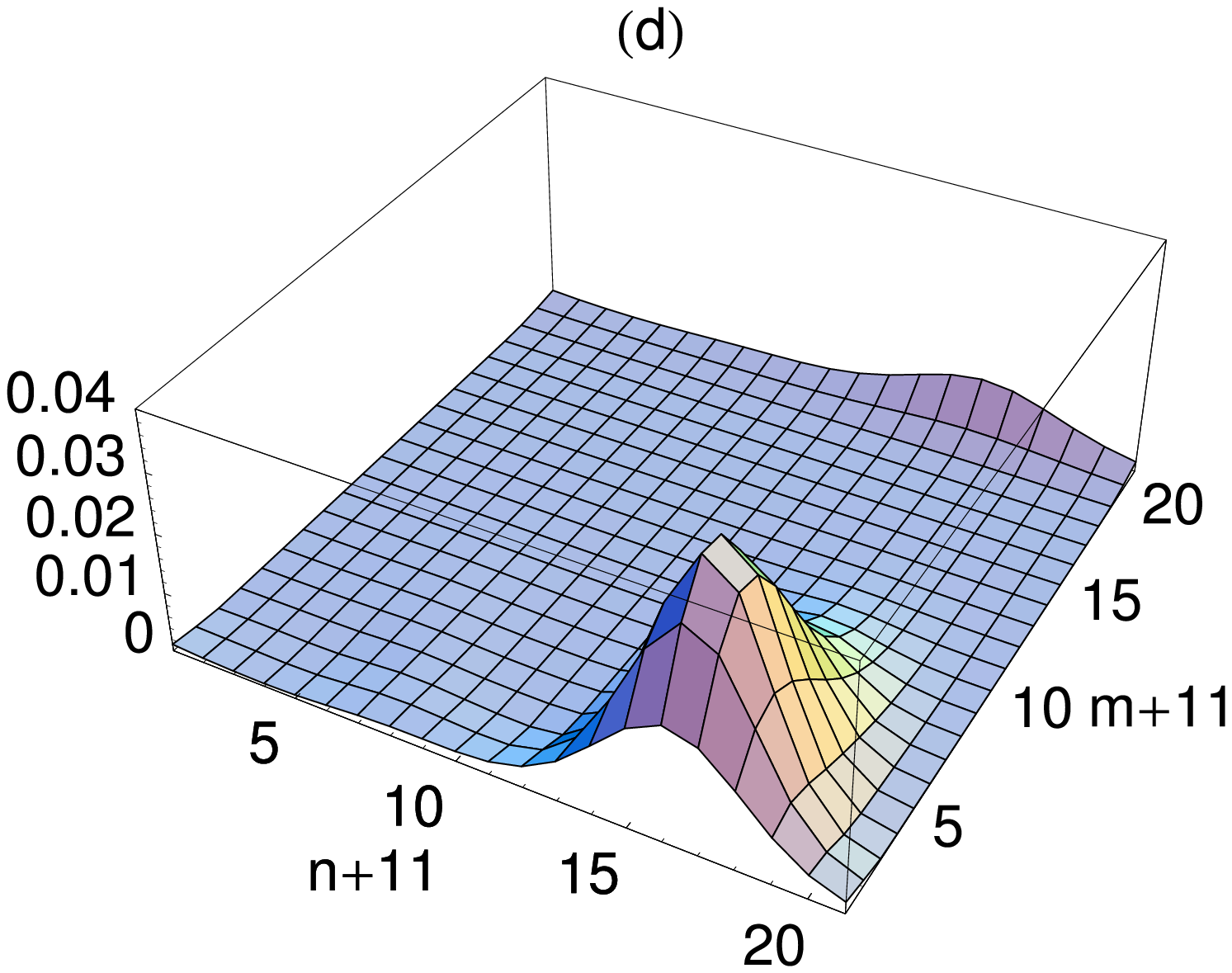}
\caption{Time evolution of $\mathcal{H}(m,n;\tau)$ for the Lipkin-Meshkov-Glick model with $\mathrm{N}_{p} = 20$ and $\chi = 1.5$ fixed,
where the labels $m$ and $n$ characterize, respectively, the dimensionless angular momentum and angle pair. These pictures show, in 
particular, how the two-body interaction term present in the Hamiltonian ${\bf H}$ affects the initial distribution $\mathcal{H}(m,n;0)$ 
for different values of dimensionless time $\tau$. We have adopted in our numerical investigations the values (a) $\tau = 0$, (b) $\tau 
\approx 6.5$, (c) $\tau \approx 15.9$, and (d) $\tau \approx 25.3$, which illustrate a representative but not complete evolution of the 
discrete Husimi function.} 
\end{figure}
After this condensed review, let us establish a sequence of steps that permits us to evaluate the time evolution of the discrete Husimi 
function for the LMG model. The first one consists in adopting the theoretical approach developed in Ref. \cite{Ruzzi3} for the 
time-dependent discrete Wigner function $\mathcal{W}(m,n;t)$ defined upon an $N^{2}$-dimensional phase space labelled by the 
angular momentum and angle pair $(m,n) \in [-\mathrm{N}_{p}/2,\mathrm{N}_{p}/2]$. Since this approach depends on the Wigner function 
evaluated at time $t=0$, the second one consists in fixing the initial state as a symmetric combination of the ground and first excited 
states following the recipe described in \cite{Pimentel}. The next and last step refers to the smoothing process given by Eq. (\ref{e3}), 
which leads us to finally obtain the desired result. Figure 1 shows the three-dimensional plots of $\mathcal{H}(m,n;\tau)$ versus
$(m,n) \in [-10,10]$ with $\mathrm{N}_{p}=20$, and $\chi = 1.5$ fixed. In the numerical investigations, we have adopted some specific
values for the dimensionless time $\tau$ in order to illustrate the effects of the two-body interaction term in the original Hamiltonian
${\bf H}$ (or, equivalently, the second term of ${\bf H}_{L}$ constituted by the operators ${\bf J}_{+}^{2}$ and ${\bf J}_{-}^{2}$) on a
given initial configuration of the finite phase space. Thus, figure 1(a) represents the discrete Husimi function $\mathcal{H}(m,n;0)$
which reflects the initial condition of the model under investigation, that is, it shows two distinct regions equally distributed in the
$21^{2}$-dimensional phase space of the particular two-level system. Figure 1(b) corresponds to a subsequent time $\tau \approx 6.5$, 
where we perceive that the probability distribution was almost totally reallocated in one side of the discrete phase space, which means 
that both the angular momentum and angle components present negative values. By its turn, figure 1(c) shows an intermediate process for
$\tau \approx 15.9$ and quite similar to that found in (a) when $\tau = 0$. Finally, figure 1(d) illustrates the migration, in $\tau 
\approx 25.3$, to a distribution of positive values of angle components in constrast to that viewed in figure 1(b). Note that, in 
particular, the quantum dynamics of this system inhibits a distribution function centered at $n=0$.  

Although the periodic pattern verified for the discrete Husimi function can be explained, in principle, through the periodic fluctuation 
of particle populations between the two energy levels, it is not clear until now its relation with the energy gap $\Delta = ( \mathrm{E}_{1} -
\mathrm{E}_{0} )/ \eps$, associated with the quasi-spin tunneling occuring in the present situation, as well as its dependence on the 
interaction parameter $\chi$. To clarify this point, some considerations on the energy spectrum related to ${\bf H}_{L}$ deserve be properly 
mentioned: (i) it is numerically obtained by just diagonalizing the matrix associated with the Hamiltonian (\ref{e5}) in the 
${\bf J}_{\fz}$-basis (for more details on the exact solutions of the LMG model, see Ref. \cite{Lipkin}); as a direct consequence of this 
result, (ii) the energy gap curve shows an explicit dependence on $\chi$, that is, as the two-body correlation strength increases, the energy 
gap decreases \cite{Lipkin,Pimentel}. Therefore, if one changes the interaction parameter $\chi$ (or the energy gap $\Delta$), the periodic 
pattern showed by $\mathcal{H}(m,n;\tau)$ will be also modified. Next, with the help of the entropy functional (\ref{e4}), we will estimate 
this quantity for the same set of parameters used in the previous figure. In this way, Figure 2 shows the plot of $\mathrm{E} \lbk \{ \lam_{i} 
\}; \tau \rbk$ versus $\tau \in [0,60]$ for $\mathrm{N}_{p} = 20$ and $\chi = 1.5$ fixed. Note that the entropy functional curve presents an
oscillatory behaviour with a well-defined periodic structure, which allows us to estimate the energy gap through the period of oscillation 
between two consecutive maximum (minimum) points. Thus, after a detailed analysis of the numerical data used in the plot of Fig. 2, we can 
conclude that $\Delta \approx 0.1784$. It is important emphasizing that this result is in perfect agreement with that obtained from the 
diagonalization process for ${\bf H}_{L}$ (in this case, the percent error estimated is $\delta \approx 0.22 \%$).

\begin{figure}[!t]
\centering
\includegraphics[width=6cm]{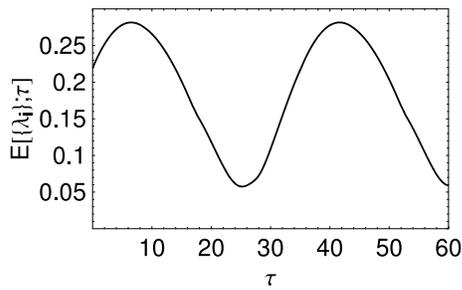}
\caption{Time evolution of $\mathrm{E} \lbk \{ \lam_{i} \}; \tau \rbk$ versus $\tau \in [0,60]$ for the same set of parameters used in
the previous figure. It is worth noticing that the maximum and minimum points of this curve correspond to the cases when the discrete
Husimi function reaches its peaks in the respective negative and positive regions of the angle sector of the finite phase space. This fact 
permits us to estimate the energy gap $\Delta = ( \mathrm{E}_{1} - \mathrm{E}_{0} )/ \eps$ from our numerical data, namely $\Delta 
\approx 0.1784$, which corroborates the value obtained through the diagonalization process of the Hamiltonian operator (\ref{e5}) in the 
${\bf J}_{\fz}$-basis ($\Delta = 0.1788$).}
\end{figure}
\begin{figure}[!t]
\centering
\includegraphics[width=6cm]{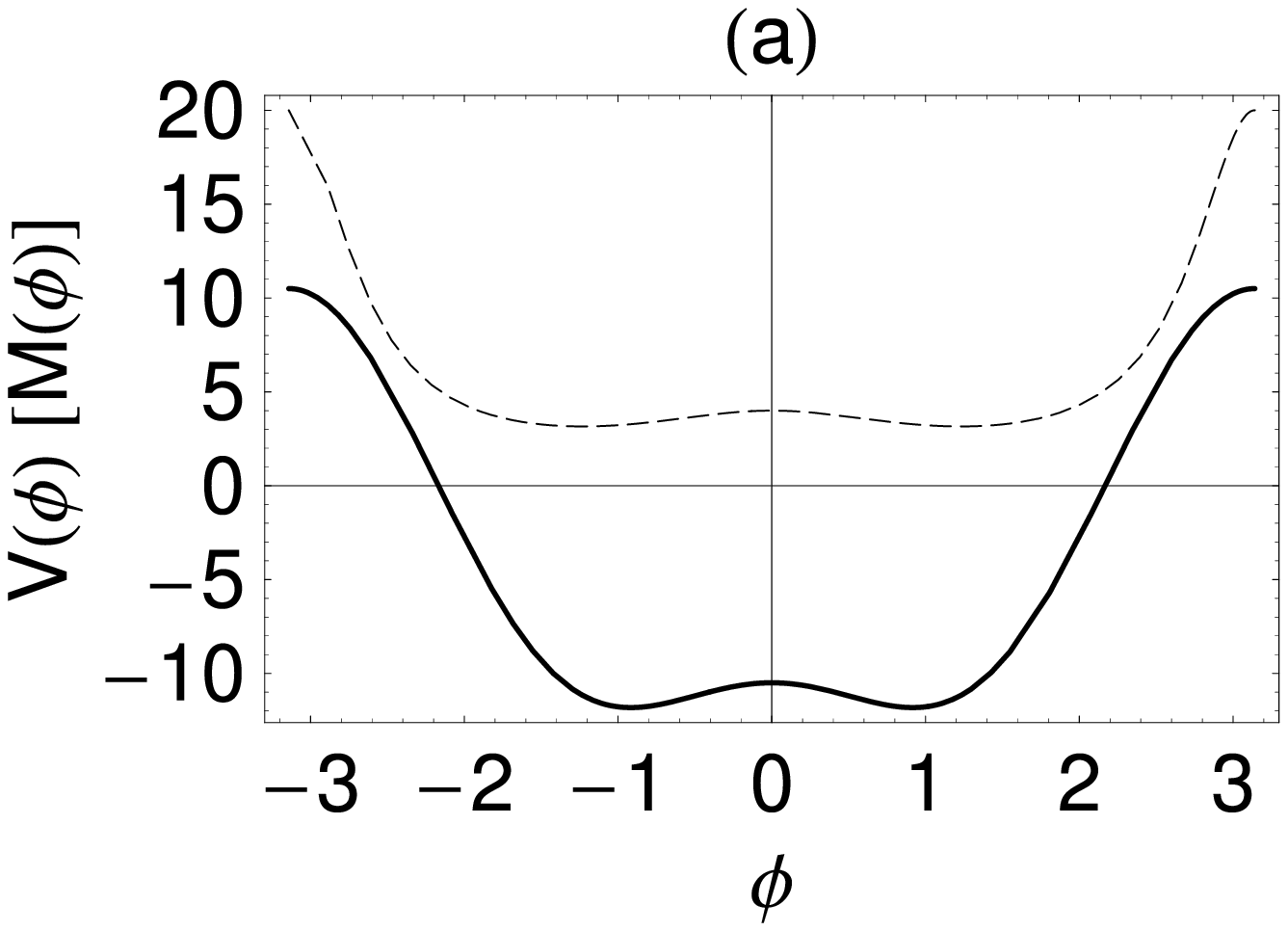}
\includegraphics[width=6cm]{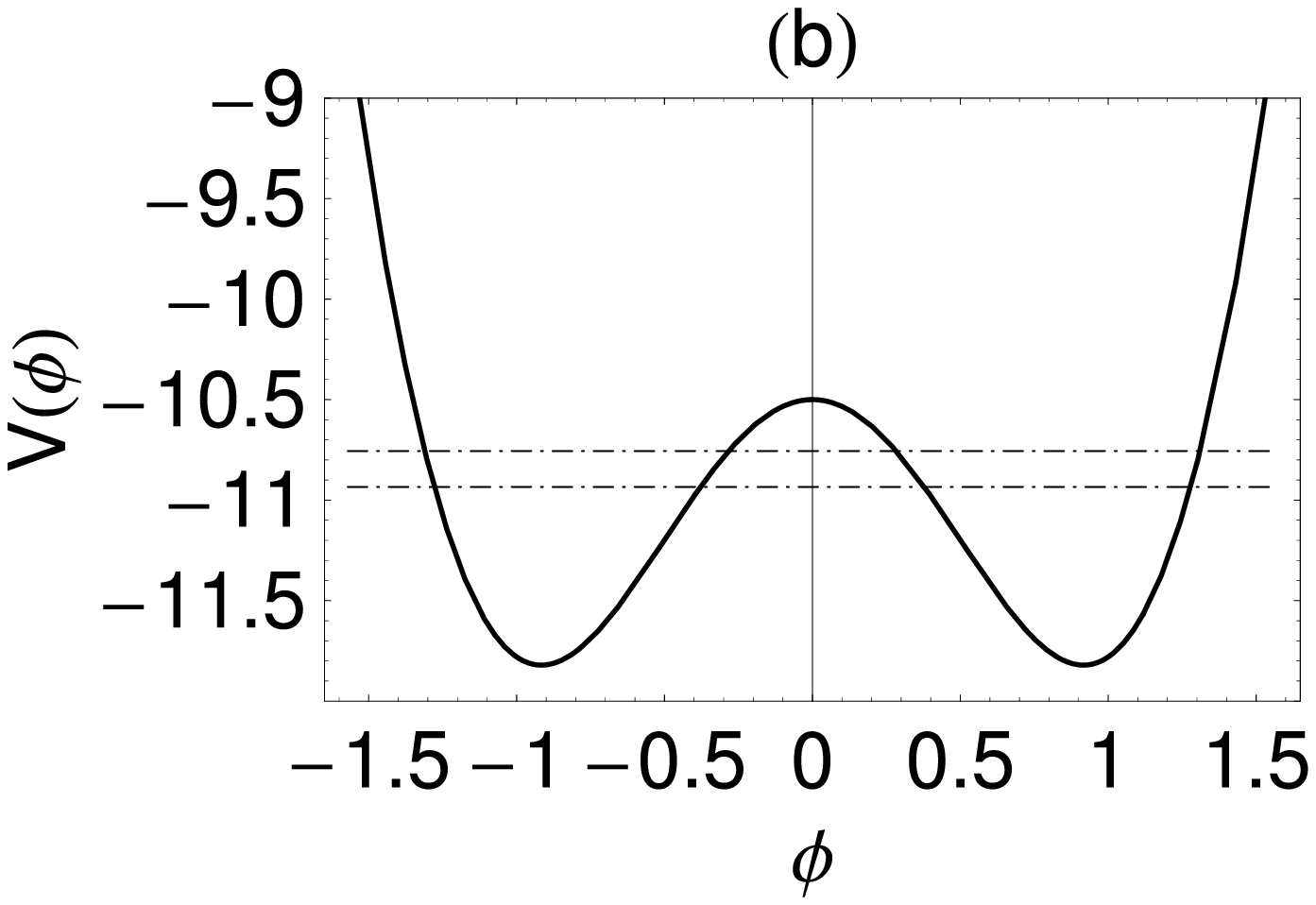}
\caption{Figure 3(a) represents the plots of $V(\phi)$ (solid line) and $M(\phi)$ (dashed line) versus $\phi \in [-\pi,\pi]$ with
$\mathrm{N}_{p}=20$ and $\chi = 1.5$ fixed. Note that both curves have similar behaviours and share complementary informations about
important quantum effects such as that appeared in figure 3(b), where now the energy levels $\mathrm{E}_{0}/ \eps$ and $\mathrm{E}_{1}/
\eps$ have also been drawn. The possibility of tunneling through the potential barrier can be blocked if the effective mass function
presents a divergent behaviour at or near the classical turning point; otherwise, the tunneling effect takes place (see both pictures
at $\phi=0$). Such description of spin tunneling, via potential and effective mass functions, can also be extended to magnetic 
molecules, such as $\mathrm{Mn}12$-acetate molecule and $\mathrm{Fe}8$ clusters.}
\end{figure}
Finally, we will present some plausible arguments that lead us to establish, under certain circumstances, a link between the oscillations 
of the discrete Husimi function, the energy gap, and the spin tunneling effect. For this intent, let us initially adopt the theoretical 
framework exposed in Ref. \cite{Galetti1}, where the potential function
\be
\lb{e6}
V(\phi) = - \frac{\mathrm{N}_{p} + 1}{2} \lbk \cos (\phi) + \frac{\chi}{2} \frac{\mathrm{N}_{p}+3}{\mathrm{N}_{p}+1} \sin^{2} (\phi) 
\rbk
\ee
and the `effective mass' function
\be
\lb{e7}
M^{-1}(\phi) = \frac{2}{\mathrm{N}_{p}-1} \lbr \cos (\phi) + \chi \lbk 1 + \sin^{2} (\phi) \rbk \rbr
\ee
were derived with details and exhaustively tested for the LMG model in the special limit $\mathrm{N}_{p} \gg 1$. Such functions, defined
in the interval $\phi \in [-\pi,\pi]$, permit us to explain, from a phenomenological point of view, the underlying behaviour of the
discrete Husimi function observed in Figure 1. For instance, figure 3(a) shows the plots of $V(\phi)$ (solid line) and $M(\phi)$ (dashed
line) versus $\phi$ for $\mathrm{N}_{p}=20$ and $\chi = 1.5$ fixed; in addition, figure 3(b) represents an amplified image of the curve 
related to $V(\phi)$, where now the lowest energy levels $\mathrm{E}_{0}/ \eps = -10.9343$ and $\mathrm{E}_{1}/ \eps = -10.7555$, extracted 
from the diagonalization process of the Hamiltonian ${\bf H}_{L}$ in the ${\bf J}_{\fz}$-basis, are also viewed (see dot-dashed lines). It 
should be stressed that distinct values of $\chi$ modify not only the energy gap but also the behaviour of the curves related to the 
potential and effective mass functions. Hence, the value here chosen for $\chi$ brings out some explicit advantages of this theoretical 
approach since the potential function presents a pronounced barrier at the origin which affects, consequently, the energy levels 
$\mathrm{E}_{0}$ and $\mathrm{E}_{1}$. Indeed, figure 3(b) consists of a paradigmatic case where tunneling effects take place in this 
quasi-spin system. Within this context, figures 1(a,c) would then correspond to a spin wavepacket equally distributed in both sides of 
the symmetric double-well potential, while figures 1(b,d) represent a wavepacket localized only on one side of the potential barrier
centered at $\phi=0$. Therefore, the time evolution of $\mathcal{H}(m,n;\tau)$ can now also be understood as a faithful representation
upon a discrete $N^{2}$-dimensional phase space (here labelled by the dimensionless angular momentum and angle pair) of spin tunneling 
processes associated with the symmetric combination of the ground and first excited states.

\section{Concluding remarks}

In what concerns the symmetry property related to the energy spectrum of the LMG Hamiltonian, let us mention some few words about the
lowest energy doublet. Numerical investigations have shown that, if one considers the energy symmetric counterpart of this doublet, the 
time evolution of the spin wavepacket will reveal a pronounced oscillatory behaviour, and this result is directly associated with the 
high energetic demand of the system in accessing these particular states. The smoothing process due to Eq. (\ref{e3}) and the subsequent 
diagonalization of the discrete Husimi function will then produce, for all $\tau \geq 0$, an entropy functional $\mathrm{E} \lbk \{ 
\lam_{i} \}; \tau \rbk$ with a well-defined periodic pattern but not related to the previously discussed tunneling effects. Therefore, 
the theoretical apparatus here exposed shows that the coherent oscillations verified in this case do not characterize the same spin 
tunneling process, although the value of the energy gap be the same.
\begin{figure}[!t]
\centering
\includegraphics[width=6cm]{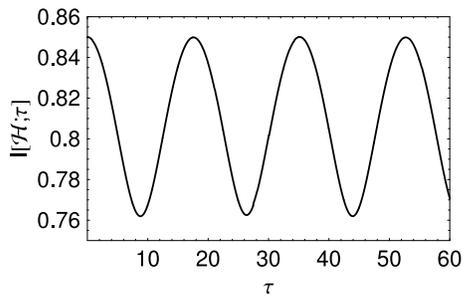}
\caption{Plot of $\mathrm{I}[\mathcal{H};\tau]$ versus $\tau \in [0,60]$ for $\mathrm{N}_{p}=20$ and $\chi = 1.5$ fixed. The oscillatory 
pattern viewed in this picture also allows us to estimate the energy gap from our numerical data, \ie, $\Delta \approx 0.1786$, with an
error of $\delta \approx 0.11 \%$.}
\end{figure}

Next, we introduce a complementary functional to that established by $\mathrm{E} \lbk \{ \lam_{i} \}; t \rbk$, which permits us to
measure, in principle, the correlation between the discrete variables of an $N^{2}$-dimensional phase space. To this end, one considers 
the mutual correlation functional \cite{Marcelo3,Vedral}
\be
\lb{e8}
\mathrm{I}[\mathcal{H};t] \coloneq \mathrm{E}[\mathcal{Q};t] + \mathrm{E}[\mathcal{R};t] - \mathrm{E}[\mathcal{H};t] \geq 0 \, ,
\ee
where $\mathrm{E}[\mathcal{H};t]$ corresponds to the time-dependent joint entropy defined in terms of the discrete Husimi function (\ref{e3}), 
with $\mathrm{E}[\mathcal{Q};t]$ and $\mathrm{E}[\mathcal{R};t]$ representing the marginal entropies which are related to the respective 
marginal distributions $\mathcal{Q}(\mu;t)$ and $\mathcal{R}(\nu;t)$ (for technical details, see Refs. \cite{Marcelo2,Marcelo3}). Thus, 
if one applies this measure to the LMG model, some interesting results can be promptly obtained. In this sense, figure 4 illustrates the 
time evolution of $\mathrm{I}[\mathcal{H};\tau]$ versus $\tau \in [0,60]$ for the same set of parameters fixed in the previous figures. From 
the numerical point of view, it is immediate to perceive that: (i) the maximum points coincide with the configurations exhibited in figures 
1(a,c) and also reflect a situation where the spin wavepacket is equally distributed in both sides of the potential barrier, which implies in 
a minimal mutual correlation (maximal uncertainty) between the angular momentum and angle pair; (ii) the minimum points describe both the
configurations illustrated in figures 1(b,d) and this result can be explained by means of a spin wavepacket localized in one of the potential 
wells, which leads us to obtain a maximal mutual correlation (minimal uncertainty) between the discrete variables $m$ and $n$; and finally, 
(iii) $\mathrm{I}[\mathcal{H};\tau]$ presents a half period if one compares with the oscillatory pattern exhibited by $\mathrm{E}[ \{ \lam_{i} 
\}; \tau]$, this fact being associated with the absence of differentiation between the cases reported in 1(b,d). 

In summary, we have developed an alternative theoretical framework for a class of physical systems described by discrete variables with 
potential applications in quantum information theory and quantum computation \cite{Nielsen}. Based on a finite-dimensional phase space 
description, this formalism was then applied to the Lipkin-Meshkov-Glick model \cite{Lipkin} whose Hamiltonian operator, although 
apparently simple, presents some remarkable physical and mathematical properties \cite{Varios}. In particular, we have shown how the 
angle-based potential approach \cite{Ruzzi3,Pimentel,Galetti2} can be used to explain qualitatively the spin tunneling effects related to 
a symmetric combination of ground and first excited states. Moreover, we have also inferred as a by-product the energy gap for this situation 
(\ie, related to the particular spin tunneling situation described here) through two different ways, and showed that both methods produce
excellent quantitative results if one compares them with the exact analogues extracted from the diagonalization process of the LMG Hamiltonian. 

Finally, it is worth mentioning that an interesting study about $\mathrm{Fe}8$ magnetic clusters in the presence of external magnetic 
fields appeared recently in \cite{Evandro}, where the theoretical apparatus for finite-dimensional phase spaces here discussed was 
extensively applied with great success in exploring the spin tunneling effects in those clusters (in particular, such quantum effects occur
for temperatures below the crossover temperature, \ie, $0.35K$ \cite{Sangregorio}). Since the $\mathrm{Fe}8$ magnetic clusters are 
characterized by a $\mathrm{J} = 10$ spin ground state \cite{note2}, some similarities with the LMG model can be shared via discrete 
phase-space approach. However, it must be stressed that such phenomenological Hamiltonian, describing the Fe8 magnetic clusters, contains 
essential experimental information concerning the anisotropies inherent to the molecule structure \cite{Caneschi}, what produces an energy 
spectrum that is essentially diferent from that obtained for the LMG model. From this theoretical approach, it was also possible to infer 
the energy gap between the ground and first excited states, namely $\Delta \approx 2.92241K$, with a percent error $\delta$ estimated in 
$1.04 \%$ (in this case, the external magnetic field applied along the easy axis of the cluster has as intensity $H_{||} = 0.11T$). 
Besides, the entropy functionals $\mathrm{E}[\mathcal{H};\tau]$ and $\mathrm{I}[\mathcal{H};\tau]$ were also employed to qualitatively 
explain how the spin tunneling effect is connected with the functional correlations between the discrete variables observed in the 
underlying phase space, corroborating, in this way, the results here presented.

\section*{Acknowledgments}

The authors thank Paulo E. M. F. Mendon\c{c}a from the University of Queensland (Australia) for providing valuable suggestions, and two 
anonymous referees for useful comments on an earlier version of this manuscript. This work has been supported by CAPES and CNPq, both 
Brazilian agencies for financial support. 


\end{document}